\documentstyle[prb,aps,twocolumn,psfig]{revtex}
\begin{document}
\draft
\title{Girvin-MacDonald-Platzman Collective Mode at  
General Filling Factors: Magneto-Roton Minimum at Half-Filled Landau Level}
\author{K. Park and J. K. Jain}
\address{Department of Physics, 104 Davey Laboratory,
The Pennsylvania State University,
University Park, Pennsylvania 16802}
\date{\today}
\maketitle
\begin{abstract}

The single mode approximation has proved useful for the excitation
spectrum at $\nu=1/3$.  We apply it to general fractions and find that it
predicts $n$ magneto-roton minima in the dispersion of the 
Girvin-MacDonald-Platzman collective mode for the fractional 
quantum Hall states at
$\nu=n/(2n+1)$, and one magneto-roton minimum for both the composite Fermi sea
and the paired composite fermion state.  Experimental relevance of the results
will be considered.

\end{abstract}


The single mode approximation (SMA) was employed 
by Girvin, MacDonald, and Platzman \cite {GMP}
(GMP) to gain insight into the excitation spectrum of the
fractional quantum Hall effect (FQHE)
at filling factor $\nu=1/3$,  where a  good
wave function for the ground state was known due to 
Laughlin \cite {Laughlin}. In analogy to Feynman's theory of superfluid
$^4$He, GMP considered a density wave excitation,  referred to below as 
the GMP collective mode, 
and predicted a ``roton" minimum in the energy dispersion, later confirmed in
exact diagonalization studies.  A sharp peak observed by Pinczuk {\em et
al.} \cite {Pinczuk1} in inelastic Raman scattering 
experiments has been interpreted as the zero wave vector limit of 
the GMP mode, and another one at lower energy \cite {Pinczuk2} as the roton
minimum.   
Now that accurate wave functions are known for all FQHE ground 
states, based on the physics of composite fermion (CF) 
\cite {Jain,Review,Review2}, 
we have investigated in this work the GMP mode for general filling fractions.
One of our motivations is to explore what insight 
the SMA provides for the 
CF Fermi sea \cite {HLR} at $\nu=1/2$, and for the fully spin 
polarized paired CF state, 
a promising candidate for the FQHE at $\nu=5/2$ \cite{Willett}.

The CF wave function for the ground state at $\nu=n/(2n+1)$ is given by 
\begin{equation}
\Phi_n^{CF}=P\Phi_1^2 \Phi_n\;.
\label{chi}
\end{equation}
$\Phi_n$ is the wave function for $n$ filled Landau levels of electrons and 
$\Phi_1^2$ attaches two vortices to each electron to convert it into
a composite fermion, hence the interpretation of $\Phi_n^{CF}$ 
as $n$ filled
Landau levels of composite fermions.  $P$ is the lowest Landau level (LL)
projection
operator.  (We will be restricting the Hilbert space to the lowest LL of
electrons throughout this work, as appropriate in the limit of very strong
magnetic fields.  Also, a strictly two-dimensional system will be considered;
finite thickness corrections, which have been
found to lower the energy gaps  by as much as
50\%,  must be incorporated in the
theory when comparing quantitatively to experiment.)
$\Phi_n^{CF}$ is an extremely good representation of the actual FQHE state at
$\nu=n/(2n+1)$ and  can be taken as exact for all practical purposes.
Just as $\Phi_n$ develops into the electron Fermi sea in the limit of
$n\rightarrow \infty$, $\Phi_n^{CF}$ 
evolves into the CF Fermi sea as $\nu=1/2$ is
approached.

Following GMP, consider the following {\em ansatz} for the excited state: 
\begin{equation}
P \rho_k \Phi_n^{CF} \equiv \overline{\rho}_k \Phi_n^{CF} 
\end{equation}
where 
$\rho_k = \sum_j \exp(-i {\bf k}.{\bf r}_j)$
is the usual density operator, and $\overline{\rho}_k$ is the 
projected density
operator.  The energy of this state is given by
\begin{equation}
\Delta E_k=
\frac{<\Phi_n^{CF}|\overline{\rho}^{\dagger}_k[V-E_0]\overline{\rho}_k|\Phi_n^{CF}>}
{<\Phi_n^{CF}|\overline{\rho}^{\dagger}_k\overline{\rho}_k|\Phi_n^{CF}>} 
=\frac{\overline{f}_k}{\overline{s}_k}
\end{equation}
$V=\frac{1}{2}\sum_{j\neq k}\frac{e^2}{|r_j-r_k|}$ 
is the Coulomb energy and $E_0$ is the energy of the ground state.
The projected static structure factor ($\overline{s}_k$)
can be obtained from the 
ordinary static structure factor by using the relation 
$\overline{s}_k=s_k-(1-e^{-k^2/2})$, and the projected oscillator 
strength ($\overline{f}_k$)
is given by the expression (with magnetic length $l_0=1$) 
\begin{equation}
\overline{f}_k=\overline{g}_k + \overline{h}_k
\end{equation}
$$\overline{g}_k=\exp[-\frac{k^2}{2}]\int_0^{\infty}  \frac{dq}{2\pi}
q V(q) \overline{s}_q [J_0(qk)-1] 
$$
$$
\overline{h}_k=2\exp[-\frac{k^2}{2}]\int \frac{d^2q}{(2\pi)^2}
V(|{\bf q}-{\bf k}|) \overline{s}_q  e^{{\bf k}.{\bf q}} 
\sin^2(\frac{{\bf k}\times{\bf q}}{2})
$$
where $V(q)=2\pi e^2/q$ is the two-dimensional 
Fourier transform of the Coulomb interaction.
Thus, $\Delta E_k$ can be calculated 
from the knowledge of the structure factor
$s_k$, which we obtain by a Fourier transformation of 
the pair distribution function, $g(r)$ \cite {Comment}.

The pair distribution function is computed numerically 
by performing Monte Carlo on systems of 50 to 60 composite fermions.
The technique for dealing with the projection operator in the CF wave function
has been described in the literature \cite {KJG,JK}.
The distance between
electrons is measured along the arc (as opposed to chord), 
which is believed to minimize the curvature effects. 
For Fourier transformation, it is useful to have an analytic
function for $g(r)$, which is obtained by expanding  
$e^{r^2/4}(g(r)-1)+e^{-r^2/4}$ in the  
power series $\sum_m C_m r^{4m+2}$ and then adjusting the coefficients 
$C_m$ to obtain the best fit \cite {GMP}.
It is necessary to keep a large number of terms in the expansion 
to ensure that all of the oscillations in $g(r)$ are captured properly, which 
in turn is crucial for obtaining the oscillations in the energy dispersion 
of the GMP mode.
At $\nu=5/11$, we keep terms up to $r^{162}$, with  a total of 41 fitting
parameters; the number of
required terms increases very rapidly as the filling factor approaches 1/2.
The fitting is done in the standard manner, 
by minimizing the chi-square function; the condition that 
the derivative of chi-square with respect to
all the fitting parameters vanish reduces to 
solving a set of linear equations for the fitting parameters. Since
we are dealing with a huge number of parameters and 
rather subtle details of raw data, we use the technique 
of ``singular value decomposition" for solving these equations, 
which fixes the 
roundoff error sensitivity of the usual normal equation
solution through Gauss-Jordan elimination.

\begin{figure}
\centerline{\psfig{figure=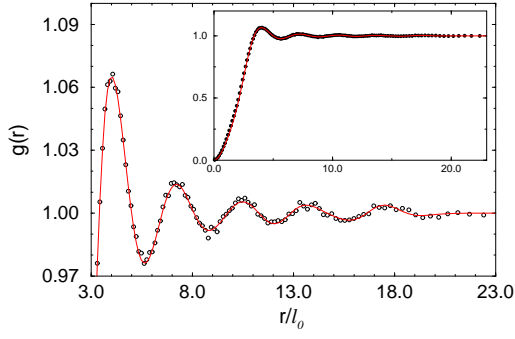,width=3.0in,angle=-90}}
\caption{The pair distribution function $g(r)$ 
for 5/11.  The points are from Monte Carlo
calculations, and the solid line is the analytical fit.  
$l_0$ denotes the magnetic length. 
\label{fig1}}
\end{figure}

For Laughlin's wave function, the pair distribution function in the 
quantum mechanical ground state is identical
to the thermal pair distribution function of a classical two-dimensional
one-component plasma (2DOCP) with logarithmic interactions.  
It must satisfy certain constraints \cite {Girvin},
which are tantamount to requiring that 
$\overline{s}_k\rightarrow \frac{(1-\nu)}{8\nu} k^4$ as $k\rightarrow 0$; in
other words, they fix the coefficients of the $k^0$, $k^2$ and the $k^4$ term
in the expansion of $\overline{s}_k$.
In terms of the 2DOCP, the absence of the
constant term is a consequence of charge neutrality, the absence of the 
$k^2$ term of perfect screening, and the coefficient
of the $k^4$ term is fixed from the compressibility sum rule. 
The first two are quite generally expected 
 for incompressible FQHE states;  given
that $\overline{f}_k\rightarrow k^4$ as $k\rightarrow 0$, a finite gap 
at small $k$ is possible only if $\overline{s}_k \sim k^4$ at small $k$.
But the coefficient of $k^4$ given above appears to be
special to Laughlin's wave function, since the other FQHE states do not
enjoy a mapping into a 2DOCP.  However,
Lopez and Fradkin \cite{LF} have argued that for any general 
incompressible state, the small $k$ properties are correctly described by 
a wave function whose modulus is given by 
$|\prod_{j<k}(z_j-z_k)|^{1/\nu}\exp[{-\sum_j\frac{|z_j|^2}{4}}]$;  
a plasma analogy on this wave function will again produce the above
coefficient of $\frac{(1-\nu)}{8\nu}$  for the $k^4$ term. 
Therefore, we fit the numerical $g(r)$ to the above power series
subject to all three constraints.  The fits are excellent as seen 
in Fig.~\ref{fig1} for $\nu=5/11$.  The goodness of the fit may 
be taken as a corroboration of the assertion made by Lopez and Fradkin with
regard to universal long distance properties of general incompressible 
fractional Hall states.  It is stressed that the fixing of the coefficient of $k^4$
in this manner is not crucial for the results below;
if we fit $g(r)$ subject to only the first two constraints, 
the dispersions of the GMP mode are affected only slightly at small $k$.

\begin{figure}
\centerline{\psfig{figure=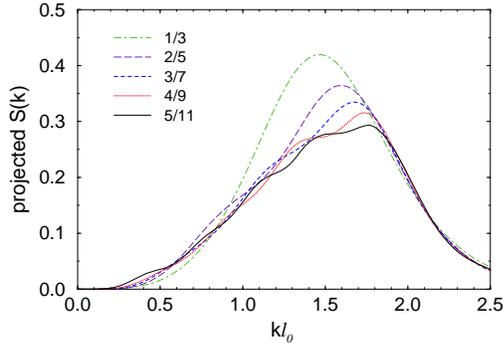,width=3.0in,angle=-90}}
\caption{Projected static structure factors $\overline{s}_k$ for the 
FQHE states at 1/3, 2/5, 3/7, 4/9, and 5/11.
\label{fig2}}
\end{figure}

\begin{figure}
\centerline{\psfig{figure=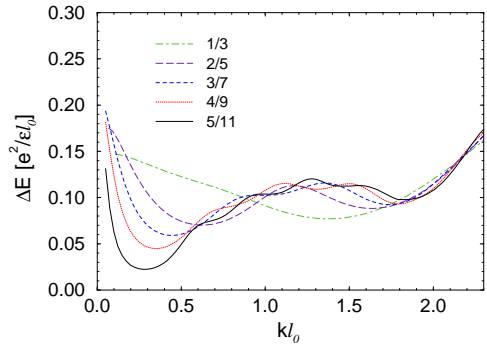,width=3.0in,angle=-90}}
\caption{Dispersions of the GMP mode at 1/3, 2/5, 3/7, 4/9, and 5/11.
The energies are in given in units of $e^2/\epsilon l_0$, where $\epsilon$ is
the dielectric constant of the background material.
\label{fig3}}
\end{figure}

\begin{figure}
\centerline{\psfig{figure=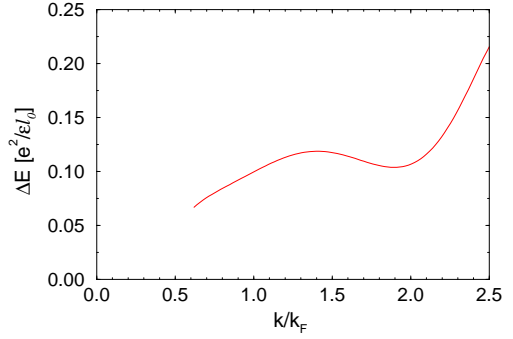,width=3.0in,angle=-90}}
\caption{The estimated dispersion of the GMP mode for the fully polarized 
composite Fermi sea at the half-filled
Landau level.  The Fermi wave vector is given by $k_F=\sqrt{4\pi\rho}$, 
$\rho$ being the electron density. 
\label{fig4}}
\end{figure}

The Fourier transformation is readily performed with the help of the 
analytic form of $g(r)$.
The projected structure factors are shown in Fig.~\ref{fig2}
for the principal $\nu=n/(2n+1)$ FQHE states at
1/3, 2/5, 3/7, 4/9, and 5/11, and the GMP-mode dispersions obtained
from them in Fig.~\ref{fig3}.  
Contrary to what one might have expected, 
the energy in the $k\rightarrow 0$ limit {\em increases} with $n$.
The small $k$ region is sensitive to the finite system size,
through the curvature of the spherical geometry.  We note that 
our $k\rightarrow 0$ limit of the energy
for $\nu=1/3$ GMP mode is in complete agreement with GMP who had employed
much bigger systems in their calculations, which 
gives us reasonable confidence that our
results are reliable even at small $kl_0$, even though we have not 
investigated systematically the particle-number ($N$) dependence of 
our results.  
There are $n$ inflection points in $\overline{s}_k$ for the state with
$n$ filled CF-LLs, producing $n$ minima in the dispersion curve.  The two
exterior minima are the strongest, with the interior minima becoming
progressively weaker with increasing $n$.  In particular, the minimum at
$kl\approx 2$ is quite robust to variations in $n$, and appears 
to  survive all the way to $\nu=1/2$ to produce a roton minimum at
$k\approx 2k_F$.  In the wave vector range $kl_0>0.5$, the dispersion is
quite insensitive to $n$ (especially if we ignore $n=1$ and 2), 
and a smoothed version, shown in Fig.~\ref{fig4}, presumably gives 
a reasonable approximation to the dispersion at $\nu=1/2$.  
No conclusions can be made for the 1/2 dispersion at smaller
$k$, due to a substantial $n$ dependence of the curves, 
and also because the wave function approach is anyway 
not expected to provide a 
reliable account of the long distance properties of the CF Fermi sea.
The significance of the roton minimum is that it is observable 
in light scattering experiments 
due to a divergence in the density of states \cite {Pinczuk}.    

An understanding of the low-energy excitations is 
intimately related to an understanding of the physics of the 
ground state, and was clarified by the CF theory: 
given that the actual FQHE ground state is 
described as the state with $n$ filled CF-LLs, 
it is natural to consider excited states in which
one composite fermion is promoted from the $n$th CF-LL to the $(n+1)$st,
producing an exciton of composite fermions.  
This provides an excellent quantitative description 
of the low-energy excitation branch at general FQHE \cite {Jain,Dev}.
In particular, the CF exciton has lower energy than the 
GMP mode;  for example, at $\nu=2/5$, the minimum energy of the GMP mode is
approximately 40\% higher than that of the CF exciton.

Even though the GMP mode is not the smallest energy excitation, it 
has a precise and important physical significance: 
It provides information about excited states that are 
connected  to the ground state by the density operator, which are
also the excitations that are probed by perturbations that couple to the
density, as for example, in light scattering 
experiments.  While the SMA is exact
when the density operator couples the ground state 
only to a {\em single} mode, or to states in a narrow range of energy,
it continues to provide the average energy (in fact, the exact first 
moment of the energy) of the density-coupled states quite generally.
Is the GMP mode observable?  In the case of the ordinary electron Fermi 
liquid at zero magnetic field, the analogous mode is the plasmon which
is sharply defined outside the single particle excitation (SPE) continuum, 
but is Landau damped inside.  However, it 
does not disappear immediately upon entering the SPE continuum; 
its line-width broadens only slowly 
as it extends deeper into the SPE region \cite {SDS}. 
We expect that the GMP mode behaves similarly, and predict 
that it will appear (say, in Raman experiments) 
as a broad peak centered at the SMA energy, possibly in addition to a  
shaper peak at lower energies coming from the CF-exciton.
Of course, the CF exciton ceases to exist for the CF sea, but the 
GMP collective mode may still be well defined and observable.

\begin{figure}
\centerline{\psfig{figure=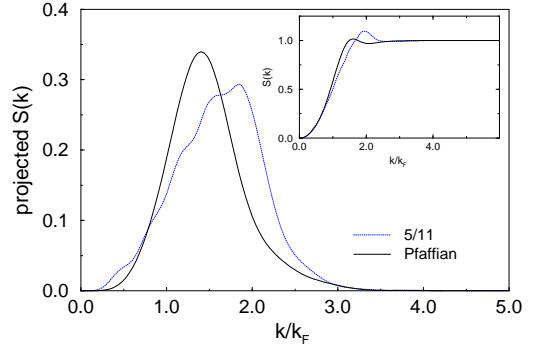,width=3.0in,angle=-90}}
\caption{Projected static structure factors $\overline{s}_k$ for the paired 
composite fermion state for $N=50$.  The full structure factor $s_k$ 
is shown in the inset.  The structure factor of the 5/11 state is also shown
for comparison. 
\label{fig5}}
\end{figure}

\begin{figure}
\centerline{\psfig{figure=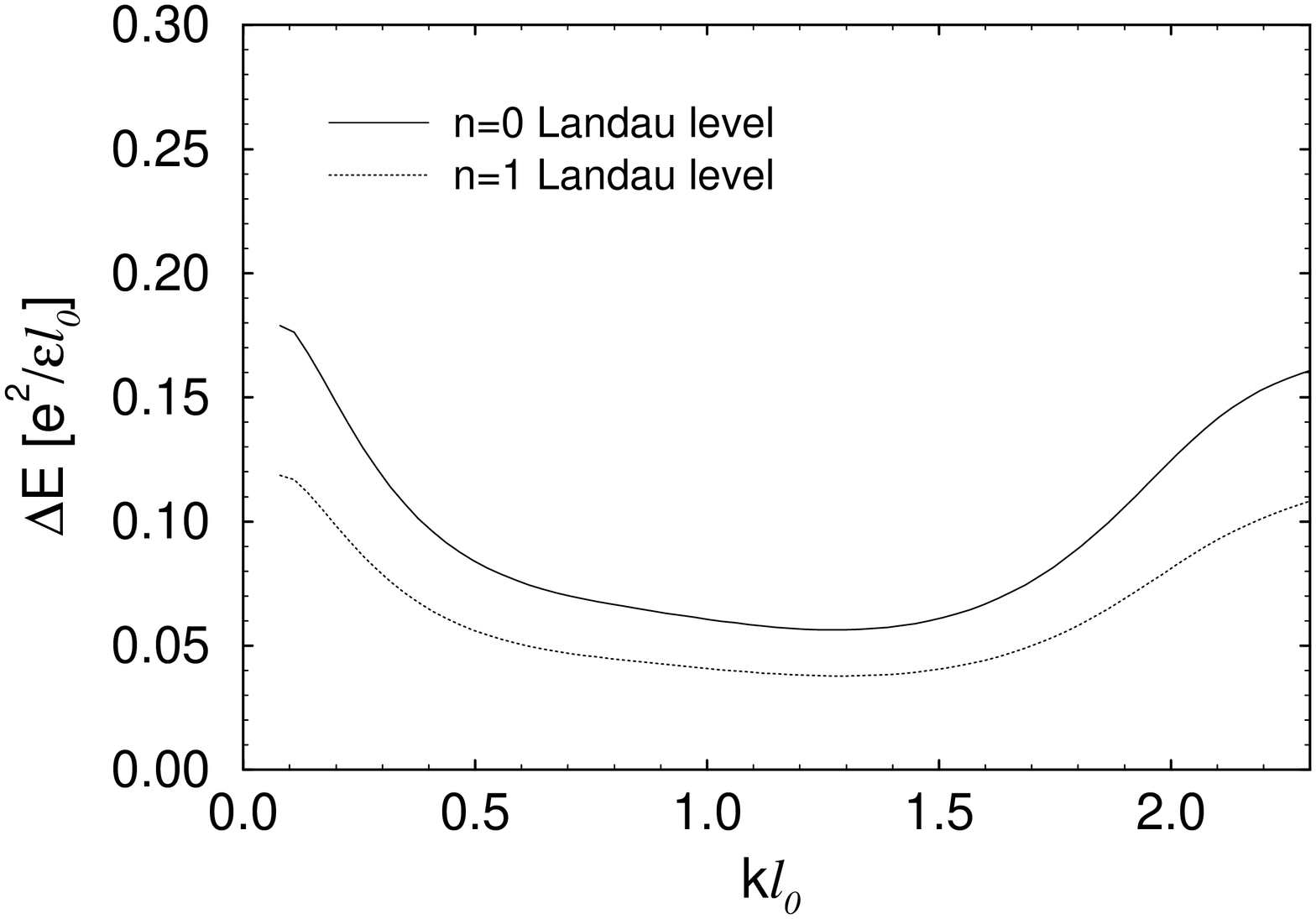,width=3.0in,angle=-90}}
\caption{Dispersion for the GMP collective mode for the paired composite 
fermion state, both for the lowest ($n=0$) and the first excited 
($n=1$) Landau levels, corresponding to $\nu=1/2$ and $\nu=5/2$, respectively.
\label{fig6}}
\end{figure}

Another interesting possibility at the half-filled Landau level is
pairing of composite fermions.  
A variational wave function for the paired CF state is 
given by \cite {Pf} 
\begin{equation}
Pf[(z_j-z_k)^{-1}]
\prod_{j<k}(z_j-z_k)^2 \exp(-\frac{1}{4}\sum_j|z_j|^2) 
\end{equation}
where $z_j=x_j+iy_j$, and, apart from an overall normalization factor, 
the Pfaffian has the form
of the real space BCS wave function: 
$Pf[(z_j-z_k)^{-1}]=A[\prod_{j=1}^{N/2}(z_{2j-1}-z_{2j})^{-1}],$
$A$ being the antisymmetrization operator. 
This state has an energy slightly higher than the CF sea in the
lowest LL, but slightly lower in the second LL \cite {PMBJ}
(i.e. at $\nu=5/2$), consistent with the experimental observation that 
a compressible state is observed at 1/2 but FQHE at 5/2.
After correcting for particle-hole symmetry, the Pfaffian state
has also been shown to have a high degree of overlap with the exact ground
state at $\nu=5/2$ in finite system studies \cite {RH}.  
All this taken together supports the view  
that the physics of the 5/2 FQHE lies in pairing of composite fermions. 
We apply the SMA to this state in order to learn about its collective
excitations.
The pair distribution function of this wave function 
has a ``shoulder" at small $r$ relative to the pair distribution function of
the CF sea \cite {PMBJ}, indicative of a real space 
pairing of composite fermions in this
wave function, and results in a structure factor, shown in Fig.~\ref{fig5}, 
peaked at a smaller wave vector than the $\overline{s}_k$ of the CF sea.  
We have computed the dispersion of the GMP mode for both the
lowest and the first excited Landau levels, appropriate for $\nu=1/2$ and
$\nu=5/2$, respectively;  the latter is obtained by using an  
effective interaction in the lowest Landau level that 
is equivalent to the Coulomb interaction in second Landau level, 
following the method outlined in Park {\em et al.} \cite 
{PMBJ}.  The resulting dispersion is shown in Fig.~\ref{fig6}; it  again 
contains a roton minimum, although much broader than for  
the CF sea.  

In summary, application of the single mode approximation to composite fermion
states has resulted in new experimentally verifiable predictions.
This work was supported in part by the
National Science Foundation under grant no. DMR-9615005,
and by NCSA Origin 2000 under grant no. DMR970015N.

\begin{thebibliography}{99}

\bibitem{GMP} S.M. Girvin, A.H. MacDonald, and P.M. Platzman, Phys. Rev.
Lett. {\bf 54}, 581 (1985); Phys. Rev. B {\bf 33}, 2481 (1986).

\bibitem{Laughlin} R.B. Laughlin, Phys. Rev. Lett. {\bf 50}, 1395 (1983).

\bibitem{Pinczuk1} A. Pinczuk {\em et al.},
Phys. Rev. Lett. {\bf 70}, 3983 (1993).

\bibitem{Pinczuk2} A. Pinczuk {\em et al.}, in {\em Proceedings of the 12th
Int. Conf. on High Magnetic Fields in the Physics of Semiconductors II}, Eds.
G. Landwehr and W. Ossau (World Scientific, 1997), p. 83. 

\bibitem{Jain}  J.K. Jain, Phys. Rev. Lett. {\bf 63}, 199 (1989);
Phys. Rev. B {\bf 41}, 7653 (1990); J.K. Jain and R.K. Kamilla in 
\cite {Review}. 

\bibitem{Review} 
{\em Composite Fermions}, edited by Olle Heinonen, (World Scientific, New
York, 1998).

\bibitem{Review2}
{\em Perspectives in Quantum Hall Effects}, edited by S. Das 
Sarma and A. Pinczuk (Wiley, New York, 1997).

\bibitem{HLR} B.I. Halperin, P.A. Lee, and N. Read, Phys. Rev. B {\bf
47}, 7312 (1993).

\bibitem{Willett} R.L. Willett {\em et al.}, Phys. Rev. Lett. {\bf 59}, 1776
(1987).

\bibitem{Comment}  Care must be exercised in obtaining the small $k$ behavior
from finite system studies 
in the spherical geometry.  For example, it has been known that at the
smallest wave vector, $k=1/R$,
where $R$ is the radius of the sphere, the projected density operator
annihilates the FQHE ground state, yielding $s_k=0$.
This, in turn, implies that $\overline{s}_k\sim k^2$ at small $k$, which is
inconsistent with the behavior $\overline{s}_k\sim k^4$, known to be valid 
in the thermodynamic limit.  This is the reason why, rather than directly 
computing $s_k$, it is more appropriate to 
obtain ${s}_k$ from $g(r)$,  after making sure that $g(r)$ satisfies certain
constraints that guarantee correct short distance behavior for
$\overline{s}_k$. 

\bibitem{KJG} R.K. Kamilla, J.K. Jain, and S.M. Girvin, Phys. Rev. B {\bf
56}, 12411 (1997).

\bibitem{JK} J.K. Jain and R.K. Kamilla, Phys. Rev. B {\bf 55}, 
R4895 (1996); Int. J. Mod. Phys. B {\bf 11}, 2621 (1997).

\bibitem{Girvin} S.M. Girvin, Phys. Rev. B {\bf 30}, 558 (1984).

\bibitem{LF} A. Lopez and E. Fradkin, Phys. Rev. Lett. {\bf 69}, 2126 (1992).

\bibitem{Pinczuk} A. Pinczuk {\em et al.}, Surf. Sci, {\bf 229}, 384 (1990);
A. Pinczuk in \cite {Review2}. 


\bibitem{Dev} G. Dev and J.K. Jain, Phys. Rev. Lett. {\bf 69}, 2843 (1992);
X.G. Wu and J.K. Jain, Phys. Rev. B {\bf 51}, 1752 (1995).

\bibitem{SDS} S. Das Sarma, private communication.


\bibitem{Pf} G. Moore and N. Read, Nucl. Phys. B {\bf 360}, 362 (1991); M.
Greiter, X.G. Wen, and F. Wilczek, Phys. Rev. Lett. {\bf 66}, 3205 (1991);
Nucl. Phys. B {\bf 374}, 567 (1992).  

\bibitem{PMBJ} K. Park, V. Melik-Alaverdian, N.E. Bonesteel, and J.K. Jain,
Phys. Rev. B {\bf 58}, R10167 (1998).

\bibitem{RH} E.H. Rezayi and F.D.M. Haldane, preprint.

\end{thebibliography}
\end{document}